\documentclass[twoside]{article}
\usepackage[sc]{mathpazo}
\usepackage[T1]{fontenc}
\usepackage{microtype}
\usepackage{amsmath}
\usepackage{mathtools}
\usepackage{amssymb}
\usepackage{graphicx}
\usepackage{color}
\usepackage[hmarginratio=1:1,top=32mm,columnsep=20pt]{geometry}
\usepackage[font=it]{caption}
\usepackage{subcaption}
\usepackage{paralist}
\usepackage{multicol}
\usepackage{mathtools, cuted}

\usepackage{lettrine}

\usepackage{abstract}

\usepackage{titlesec}
\renewcommand\thesection{\Roman{section}}
\titleformat{\section}[block]{\large\scshape\centering}{\thesection.}{1em}{}

\usepackage{fancyhdr}
\usepackage{hyperref}

\title{\vspace{-15mm}%
	\fontsize{24pt}{10pt}\selectfont
	\textbf{An Entropic Model of Gaia}
	}	
\author{%
	\large
	\textsc{R. Arthur${}^{*1}$ and A. Nicholson${}^{2\sharp}$} \\[2mm]
	\normalsize	{\it ${}^1$University of Exeter. }\\
	\normalsize	{\it ${}^2$University of Exeter, Earth System Science. } \\
	\normalsize	${}^*$\href{mailto:rudy.d.arthur@gmail.com}{rudy.d.arthur@gmail.com} ${}^\sharp$\href{mailto:arwen.e.nicholson@gmail.com}{arwen.e.nicholson@gmail.com}
	\vspace{-5mm}
	}
\date{}

\begin{document}

\maketitle
\thispagestyle{fancy}

\begin{abstract}
\noindent 
We modify the Tangled Nature Model of Christensen et. al. so that the agents affect the 
carrying capacity. This leads to a model of species-environment co-evolution where the system tends to have
a larger carrying capacity with life than without. We discuss the model as an example of an entropic hierarchy and
some implications for Gaia theory.
\end{abstract}

\section{Introduction}

%describe model
The first goal of this paper is to show how the Logistic growth model is intimately related to the 
Tangled Nature Model(TNM) \cite{Christensen:2002}. The TNM has
been extensively explored and elaborated by Jensen and others \cite{Anderson:2004}, \cite{Lawson:2006},
\cite{Laird:2006}, \cite{Laird:2007}. The TNM was originally developed to focus on co-evolution 
and to study the time development of macroscopic ecological observables, 
such as species diversity and total population, in co-evolutionary systems. The characteristic macroscopic features 
of a single Tangled Nature
history are long periods of stability separated by abrupt, spontaneous, transitions. These stable periods,
called quasi-evolutionary stable states (q-ESS) in the literature, are characterised by a small group of symbiotic 
`core' species which account for most of the population and a `cloud' of mutants with random, positive or negative
interactions with each other \cite{Becker:2013}. The core and cloud dynamics are crucial to understand the model and
we will discuss them extensively in what follows.

In this work we suggest a generalisation of the TNM - that the single parameter representing the carrying capacity 
becomes a function of the type and population of other species present in the system. 
We have three terms contributing to the fitness of a species $a$ in this extended model: 
\begin{itemize}
\item A term modelling the direct effect of individual $b$ on $a$ (e.g. $b$ eats $a$)
\item A term modelling the effect of individual $b$ on the physical environment of $a$ (e.g. $b$ nests at the same
sites as $a$)
\item A term modelling an interaction between $a$ and $b$ whose strength is proportional to the population of $b$.
\end{itemize}
This third term accounts for situations in nature where the by-products of one species can have 
effects on other species and their ability to reproduce. This brings us to our main aim:
connecting this model to ideas about life's interaction with the earth and the body of work that is Gaia theory \cite{Lovelock:1972},
\cite{Lovelock:1974}, \cite{Lenton:1998}. 

%describe Gaia
Gaia theory remains somewhat controversial (see e.g. \cite{Doolittle:1981}, \cite{Dawkins:1983} or more recently
\cite{Tyrrell:2013}). One part, which is more or less accepted, is that living organisms interact with and influence
their inorganic environment in what can be called species-environment co-evolution.
More controversial are the assertions that life maintains habitability, e.g by acting
as a thermostat to keep surface temperatures within tolerable limits. Even more controversial is the idea that
life is optimising the earth to make it more habitable. The principal
objections have been based on the idea that cheaters, who benefit from the
improved environment without contributing, would quickly out compete the other species, collapsing the system.
Our model addresses just this point: we have many individuals of different species which are more likely to 
reproduce if they have high fitness - given by the sum of inter-species and species-environment interactions.
We find many situations where new species exploit the environment at catastrophic cost to
the extant species and ultimately themselves. However we will find that, while the habitability of a single
system may fluctuate up and down, across multiple systems there is a tendency for stability to
increase and for life to improve habitability. We will then discuss the mechanisms causing this, which are largely entropic.

Section \ref{sec:connect} describes the connection between the TNM and Logistic model, and can
be skipped by readers only interested in Gaia theory. 
We introduce our new model in section \ref{sec:newmodel}, describe how we perform simulations in section \ref{sec:simulations}
and show averages across multiple histories in section \ref{sec:results}. Our main discussion of how
the evolutionary dynamics leads to Gaia (improved stability and improved habitability
as a consequence of life) is given in section \ref{sec:discussion} and we conclude in section \ref{sec:conclusion}.

\section{Tangled Nature and the Logistic Model} \label{sec:connect}

The fundamental quantity in the TNM is the reproduction probability 
\begin{align}\label{eqn:repro}
p(f_i) = \frac{1}{1 + e^{-f_i}}
\end{align} 
a sigmoid function which takes the fitness of species $i$, $f_i$, and returns a number in $(0,1)$ that is taken to be
the probability for an individual of that species to reproduce. The TNM update step
consists of choosing an individual, reproducing with this probability
and then removing an individual with probability $p_{k}$ (constant for all species).
We set the mutation rate to $0$ for simplicity, though later when we come to do TNM simulations
we will have non-zero mutation rates. We can redefine $p(f_i)$ by adding a constant, $A$, to raise the threshold 
fitness below which reproduction is very unlikely (or we can imagine shifting the whole fitness landscape up or down by
a constant amount)
\begin{align}\label{eqn:repro2}
p(f_i) = \frac{1}{1 + e^{-f_i+A}}
\end{align} 
For species $i$, with population $N_i$, the average number of reproduction events is
$N_i p(f_i)$ and the average number of deaths is $N_i p_k$, thus the rate of change of population of species $i$ is roughly
\begin{align}
\frac{dN_i}{dt} = N_i \left( p(f_i) - p_k \right)
\end{align}
For values of $f_i \simeq A$ the logistic function \ref{eqn:repro2} is approximately a straight line
\begin{align}
p(f_i) \sim \frac{1}{2} \left( 1 - \frac{A}{2} + \frac{f_i}{2} \right) 
\end{align} 
Since $A$ is arbitrary, let $A = 2 + 4 p_k$ then,
\begin{align}
\frac{dN_i}{dt} \sim N_i \frac{f_i}{4}
\end{align}

For the TNM with the fitness function is chosen to be \cite{Christensen:2002}
\begin{align}\label{eqn:fitness}
f_i &= \sum_j J_{ij} n_j - \mu N
\end{align}
In this and all other sums, if unspecified, the index ranges over all extant species.
For simplicity we will absorb the factor $1/4$ into the definitions of $J_{ij}$ and $\mu$ so that
we are left with the equation 
\begin{align}\label{eqn:logisticTNM}
\frac{dN_i}{dt} = N_i \left( \sum_j J_{ij} n_j - \mu N \right)
\end{align}
for the average change in the population of $N_i$.

The Verhulst or Logistic growth model is much simpler. It is a differential equation
\begin{align}
\frac{dN_i}{dt} &= r_i N_i \left( 1 - \frac{N_i}{K_i} \right) = N_i \left( r_i - \mu_i N_i \right) 
\end{align}
which describes a single species $i$, with population $N_i$, growing with a resource constraint.
$K_i$ is the carrying capacity, equal to the population at equilibrium $\frac{dN_i}{dt} = 0$,
and the second form is a simple rewriting of the first with $\mu_i = \frac{r_i}{K_i}$. 

The idea of the Tangled Nature Model, and co-evolution in general, is that the growth rate of a single species is dependent 
on the other species present in the ecosystem i.e. $r_i \rightarrow r_i(\vec{n})$ where
\begin{align*}
\vec{n}_i = \frac{N_i}{N}, \qquad N = \sum_j N_j \nonumber
\end{align*}
We can Taylor expand $r_i$ around the equilibrium $\vec{n} = 0$.
\begin{align}
r_i( \vec{n} ) &= r_i( 0 ) + \sum_j \frac{dr_i}{dn_j}(0) n_j + \ldots \\ \nonumber
r_i( \vec{n} ) &\simeq \sum_j J_{ij} n_j \nonumber
\end{align}
truncating at the linear term and defining $J_{ij} = \frac{dr_i}{dn_j}$.  
We set $r_i( 0 ) = 0$ so that no species can grow 
independently of all others.
This expansion is accurate when no single species makes up the majority of the population: $n_j \ll 1$
for all $j$. Substituting we get
\begin{align}\label{eqn:logistic}
\frac{dN_i}{dt} &=  N_i \left( \sum_j J_{ij} n_j - \mu_i N_i \right) 
\end{align}
This is the Logistic growth model in the case where the growth rate is no longer
intrinsic but depends the other species present in the ecosystem.

Comparing equations \ref{eqn:logisticTNM} and \ref{eqn:logistic} we see that the average growth rate of the 
TNM with no mutation and the Logistic growth model, where growth rate is a linear function of interspecies interactions,
are very similar. The difference is in the damping term, $\mu N$ for the TNM versus $\mu_i N_i$ for the
Logistic model. In the Logistic model a species' growth is only constrained by its own
population, while in the TNM a species' growth is constrained by the total number of individuals in the system.
Either case may be more or less realistic depending on the ecosystem under consideration e.g. for multiple
bacterial cultures growing on the same medium {\it in vitro} $\mu N$ may be appropriate, or for an ecosystem
where a single bird species competes for nesting sites $\mu_i N_i$ may be better. By considering a generalisation
of the Logistic model we can allow for these two scenarios.

\section{Species-Environment Interactions in the Tangled Nature Model} \label{sec:newmodel}

The N-species competitive Lotka-Volterra equations  (see e.g. \cite{Kondoh:2003}, \cite{Ackland:2004}) are,
\begin{align}\label{eqn:cLK}
\frac{dN_i}{dt} = N_i \left( r_i - \sum_{j} \mu_{ij} N_j \right) 
\end{align}
These equations generalise the Logistic model by making the damping term a weighted sum of the effects of each species
on $i$. We can recover the standard TNM form by putting $\mu_{ij} = \mu = \text{constant}$ or get the
Logistic form by putting $\mu_{ij} = \mu_i \delta_{ij}$. $\mu_{ij}$ represents
the effect of $j$ on the carrying capacity of the system for individuals of species $i$.\footnote{
Equation \ref{eqn:cLK} is often written using $\mu_{ij} = \frac{r_i \alpha_{ij}}{K_i}$. We use $\mu_{ij}$ to 
be closer to the standard notation for the TNM.}

Motivated by equation \ref{eqn:cLK} we generalise the TNM fitness function to be
\begin{align}
f_i &= \sum_j J_{ij} n_j - \sum_j \mu_{ij} N_j
\end{align}
Just as we did with the
growth rate we can expand the damping term as a function of $\vec{n}$:
\begin{align*}
\mu_{ij}( \vec{n} ) &= \mu_{ij}(0) + \sum_{k} \frac{d \mu_{ij} }{dn_k} n_k + \ldots \\ \nonumber 
\mu_{ij}( \vec{n} ) &\simeq \mu_{ij}(0) + \sum_{k} \eta_{ijk} n_k \nonumber 
\end{align*}
Now the fitness is
\begin{align}
f_i &= \sum_j n_j J_{ij} - \sum_j \left( \sum_{k} \eta_{ijk} n_k  + \mu_{ij}(0) \right) N_j 
= \sum_j n_j J_{ij} - \sum_j \mu_{ij}^{eff} N_j \nonumber
\end{align}
Where the effect of $j$ on the habitability for $i$ is now
$\mu_{ij}^{eff} = \sum_{k} \eta_{ijk} n_k  + \mu_{ij}(0)$. Alternatively we can write the fitness as
\begin{align*}
f_i &= \sum_j n_j \left( J_{ij} - \sum_{k} \eta_{ijk} N_k \right) - \sum_j \mu_{ij}(0) N_j =  \sum_j n_j J_{ij}^{eff} - \sum_j \mu_{ij}(0) N_j \nonumber
\end{align*}
With $J_{ij}^{eff} = J_{ij} - \sum_{k} \eta_{ijk} N_k$. This is an effective interaction
between $i$ and $j$ that depends on the population of all extant species. 
The term $\eta_{ijk} N_k$
in $J_{ij}^{eff}$ means that as the population of species $k$ increases, its effect on species $i$
may go from positive to negative, become more positive, become more negative or go from 
negative to positive - depending on the signs of $\eta_{ijk}$ and $J_{ij}$. There are cases like this in nature: 
\begin{itemize}
\item Small numbers of algae are beneficial food sources for fish but algal blooms can be deadly.
\item Small numbers of gut bacteria provide useful digestive functions for ruminants but large numbers
of fermenters cause toxic by-products like ammonia.
\item 2.3 billion years ago small numbers of photosynthesising bacteria may have been
useful food sources for other species (or at least not directly harmful) until the respiration of large numbers of them
caused a build up of oxygen in the atmosphere, triggering a mass extinction: the Great Oxidation Event.
\item In a similar way, ruminants may be benign members of an ecology but in very large numbers their emissions
can alter global temperatures, which can affect even very distant species.
\item Photosynthesisers also provide a positive example. They have
minimal direct interaction with carnivores, but provide the oxygen they breath.
\end{itemize}
Motivated by examples like these we put $\eta_{ijk} = \delta_{jk} M_{ij}$ and we also set
$
\mu_{ij}(0) = \mu = \text{constant}
$
, to limit the total population as in the standard TNM, by having every species contribute 
equally to a global damping term. Now
\begin{align}\label{eqn:fitness2}
f_i &= \sum_j n_j J_{ij} - \mu N - \sum_j n_j M_{ij} N_j
\end{align}
We also set $M_{ii} = 0$, so that all of a species' effect on its own environment is given by 
its contribution to the $-\mu N$ term. 

\subsection{Measuring Habitability}
In equilibrium, with no mutations or death, the average population of a species follows
\begin{align*}
\frac{dN_i}{dt} &= N_i \sum_{j} J_{ij} n_j - N_i \sum_j M_{ij} n_j N_j - N_i \mu N \\ \nonumber
&= N_i \left( (r_i - e_i) - \mu N \right) \nonumber
\end{align*}
where we defined 
$r_i = \sum_{j} J_{ij} n_j$ and $e_i = \sum_j M_{ij} n_j N_j$. The net growth rate of species $i$ is therefore 
$r_i - e_i$. Summing over all species and shuffling the terms gives,
\begin{align}
\sum_i \frac{dN_i}{dt} = \frac{dN}{dt} &= \sum_i N_i \sum_{j} J_{ij} n_j - \mu N^2- \sum_i N_i \sum_j M_{ij} n_j N_j \\ \nonumber
	&= N \sum_i n_i \sum_{j} J_{ij} n_j - \mu N^2 - N^2 \sum_i n_i \sum_j M_{ij} n_j^2 \\ \nonumber
	&= r N - (\mu-E) N^2  
\end{align}
where we put $r=\sum_i n_i \sum_{j} J_{ij} n_j$ and $E = -\sum_i n_i \sum_j M_{ij} n_j^2$. This
is a logistic equation with carrying capacity at equilibrium $\frac{r}{\mu-E}$. 
Thus increasing $E$ increases the carrying capacity of the system. The term $-\mu N^2$ represents a limit to growth that is
independent of which species are actually realised, while the term $E N^2$ limits or encourages growth 
depending on which species are realised. We will refer to $E$ as the `habitability'.
The situation is roughly analogous to how a real system can have physical constraints
which cannot be altered by life, $\mu$, (volcanic activity, solar flux) and constraints which can, $E$,
(soil composition, oxygen abundance, $CO_2$ insulation). Gaia theory deals with this second class of constraints 
and so $E$ will be of interest to us.

\section{Simulations}\label{sec:simulations}
We are interested in how the terms making up the TNM fitness in equation \ref{eqn:fitness2} balance when we evolve with a fixed
mutation rate. More details about performing TNM simulations can be found in references \cite{Christensen:2002}
and \cite{Arthur:2016}. To populate $M_{ij}$ and $J_{ij}$ every element is set equal to the product of two random
Gaussian numbers (mean zero and variance one) and a constant. For $J_{ij}$ the constant is
$c=100$, for $M_{ij}$ we multiply by $\sigma$ where $\sigma=0$ corresponds to the standard TNM.
A fraction $\theta = 0.75$ of the elements of $J_{ij}$ are set to zero but we do not set any of the elements of $M_{ij}$ to zero.  
We label individuals by a string of twenty 0s and 1s, which we call the genome. A species is a group
of $N_i$ individuals with the same genome $i$. With each reproduction event we make two copies of the parent with probability
$p_{mut} = 0.01$ to mutate each base (flip a 0 to a 1 or vice versa).
We set $\mu = 0.1$ and the death rate $p_k = 0.2$. The basic TNM update consists of
\begin{itemize}
\item Choosing an individual of species $i$ and reproducing that individual with probability 
$p(f_i) = \frac{1}{1 + e^{-f_i}}$ and mutation rate $p_{mut}$.
\item Choosing an individual and killing it with probability $p_k$.
\end{itemize}
A sequence of $\frac{N}{p_k}$ updates makes one `generation', which is the timescale we use for the model. 
All simulations will be run for 200000 generations and we will perform 1000 runs using different random seeds
and average over all runs.

We set $A=0$, as is standard in the TNM literature. This means species with zero fitness 
have $p(0) = \frac{1}{2}$, and so will often reproduce. This is useful for starting the simulations,
allowing an ecology to be generated spontaneously from a single starting species. We note that
with the additional population dependent interaction the global damping $-\mu N$ is not enough
to always ensure a bounded population: consider two species $a$ and $b$ with
\begin{align*}
N_a = N_b = \bar{N}, \quad J_{ab} = J_{ba} = 0, \quad M_{ab} = M_{ba} = -m, \quad m>0
\end{align*}
Then
\begin{align}\label{eqn:explosion}
f_a = f_b = \bar{N} \left( \frac{1}{2} m  - 2 \mu \right) 
\end{align}
Which is positive and increasing with $\bar{N}$ for $m > 4 \mu$, meaning the reproduction
probability tends to 1 and the population exponentially increases. We could modify the model
in a variety of ways to avoid this situation e.g. add another damping term so that $f_a' \rightarrow f_a - \mu_2 N^2$
with $\mu_2 \ll \mu$. To avoid introducing more parameters
we work with values of $\sigma$ such that this situation occurs rarely in practice ($\sim$1\% of
simulation runs), and we leave out runs where it occurs from our final statistics. This gives a slight bias away from
large negative values of $M_{ab}$ but does not affect any of our conclusions. 

\section{Results}\label{sec:results}

\begin{figure}
    \centering
       \begin{subfigure}[t]{0.4\textwidth}
        \includegraphics[width=\textwidth]{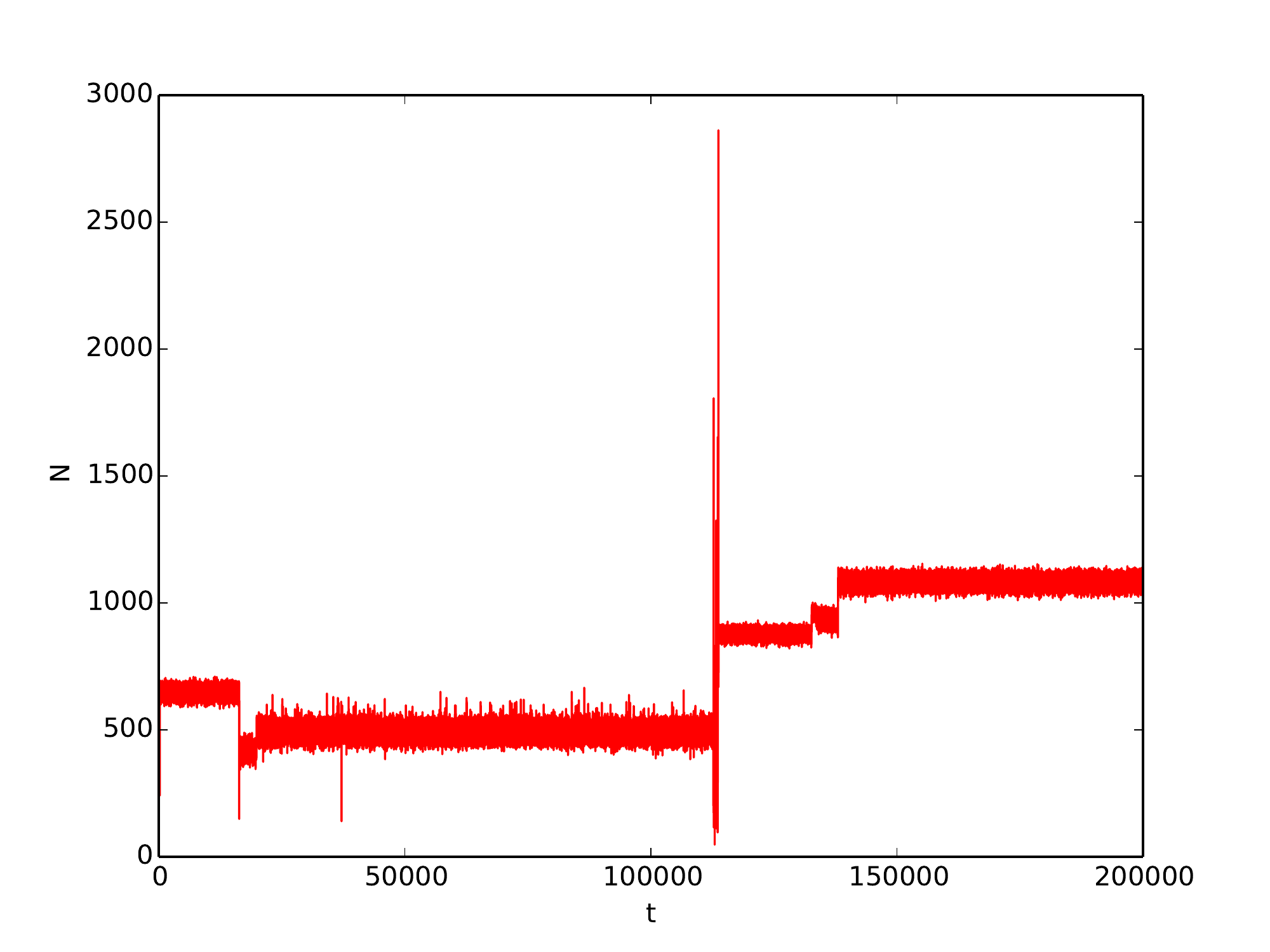}
        \caption{ Total population, $N$, for a single run of $2 \times 10^5$ generations of the TNM with $\sigma=0.1$. The sudden
        discontinuities are quakes, which signify the end of one stable regime and the start of another. }
        \label{fig:run}
    \end{subfigure}  \qquad
    \begin{subfigure}[t]{0.4\textwidth}
        \includegraphics[width=\textwidth]{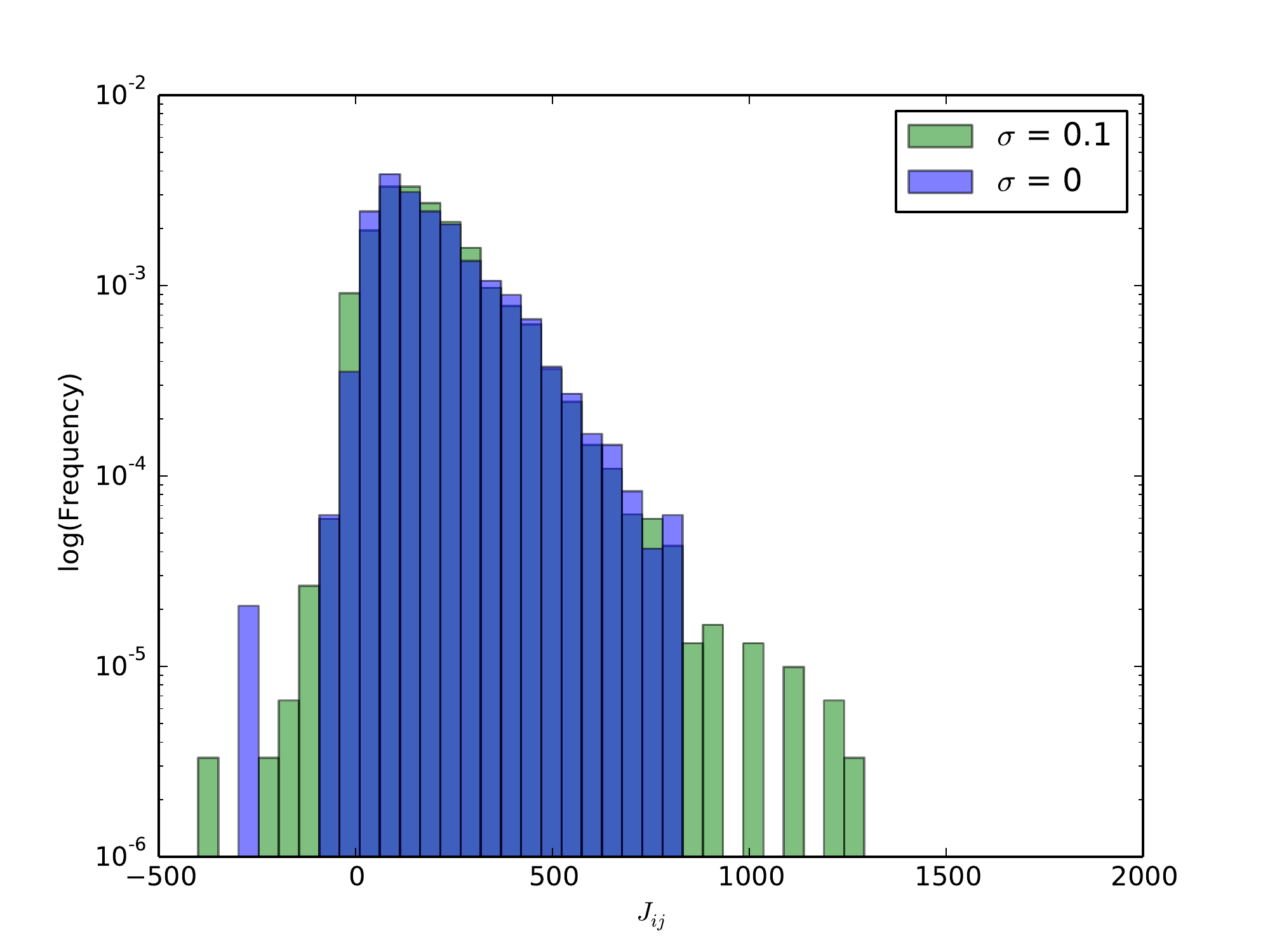}
        \caption{ Distribution of $J_{ij}$ values realised by the core in 1000 TNM runs after $2 \times 10^5$ generations. }
        \label{fig:J}
    \end{subfigure}  
    
    \begin{subfigure}[t]{0.4\textwidth}
        \includegraphics[width=\textwidth]{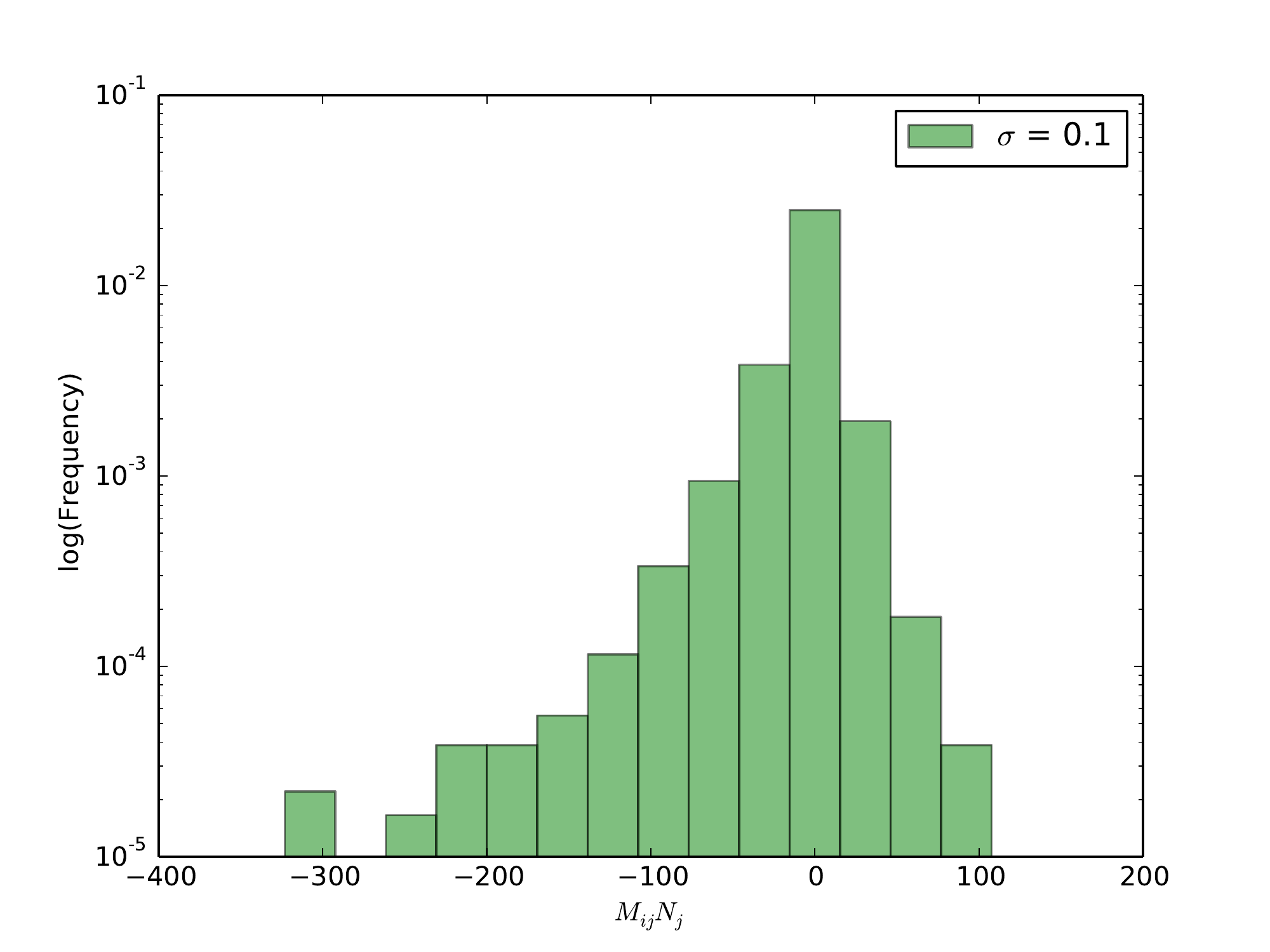}
        \caption{ Distribution of $M_{ij} N_j$ values realised by the core in 1000 TNM runs after $2 \times 10^5$ generations. }
        \label{fig:mun}
    \end{subfigure} \qquad
        \begin{subfigure}[t]{0.4\textwidth}
        \includegraphics[width=\textwidth]{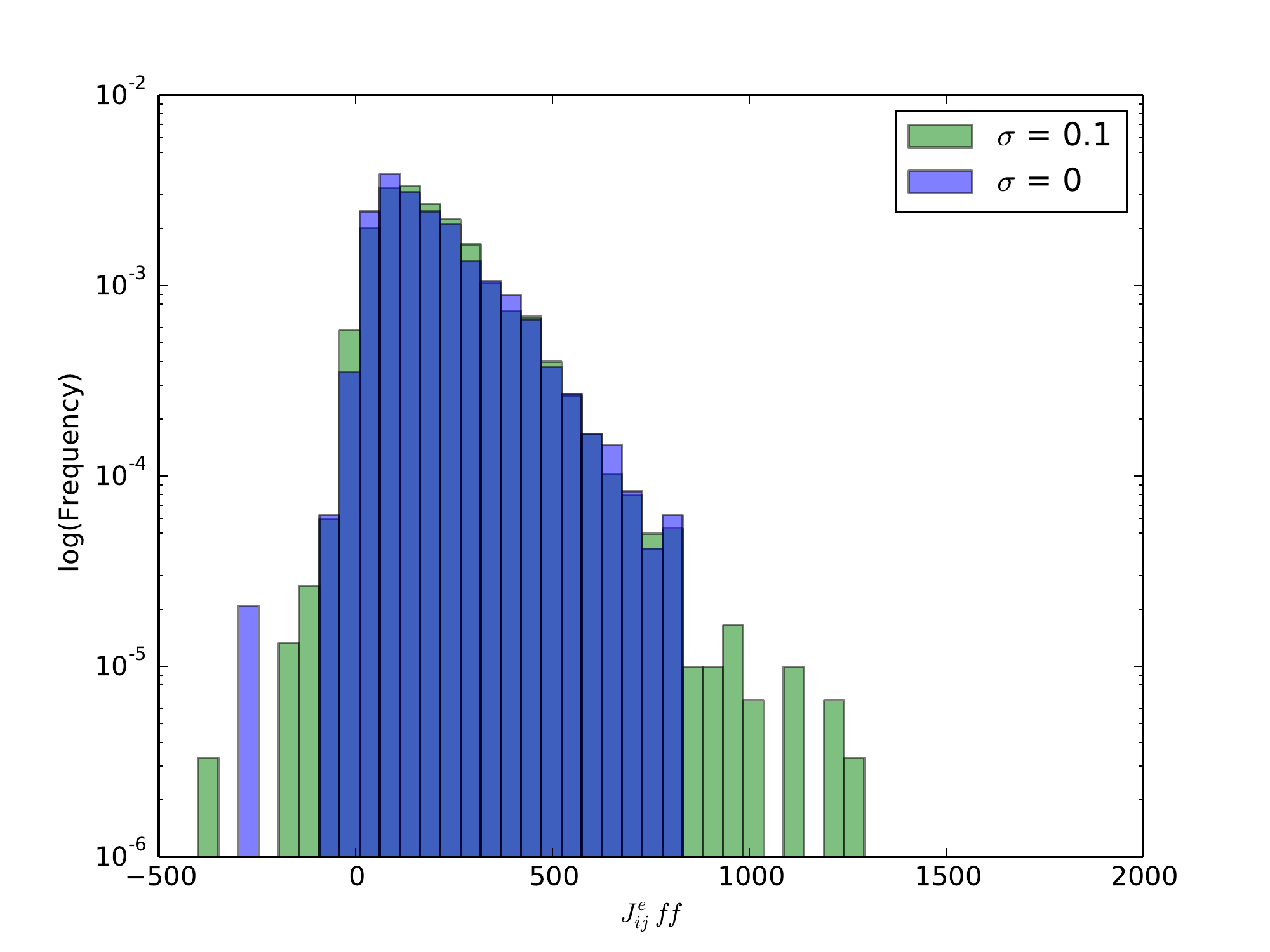}
        \caption{ Distribution of $J_{ij}^{eff}$ values realised by the core in 1000 TNM runs after $2 \times 10^5$ generations. }
        \label{fig:jeff}
    \end{subfigure} 
    \caption{(a) A single TNM run and (b-d) Histograms of the core-core coupling values. $\sigma=0$ is the standard TNM
    and $\sigma = 0.1$ is the generalisation to nonzero $M_{ij}$, where individuals affect habitability.}\label{fig:hist}
\end{figure}

We find that the standard TNM phenomenology is robust, see figure \ref{fig:run}. Most of the time the system is in
a stable state with periodic disruptions occurring spontaneously: a punctuated equilibrium. There is a small group 
of {\it core} species with mutually positive interactions which account for the majority of the population, 
and a {\it cloud} of other species with small populations that 
account for the majority of the diversity. A core species is operationally defined as one whose population is greater than
5\% of the population of the most populous species \cite{Becker:2013}.
We plot the distribution of $J_{ij}$, figure \ref{fig:J}, $M_{ij} N_j$, figure \ref{fig:mun}, and $J_{ij}^{eff}$ , 
figure \ref{fig:jeff}, realised in core-core interactions. 
The case $\sigma = 0$ corresponds to the standard TNM. As in the standard TNM the interactions $J_{ij}$
and $J_{ij}^{eff}$ in the core are positive. The distribution of $M_{ij} N_j$ is strongly skewed to the left.
This means core groups with negative $M_{ij} N_j$, improving the environment, are favoured.

\begin{figure}
    \centering
       \begin{subfigure}[t]{0.4\textwidth}
        \includegraphics[width=\textwidth]{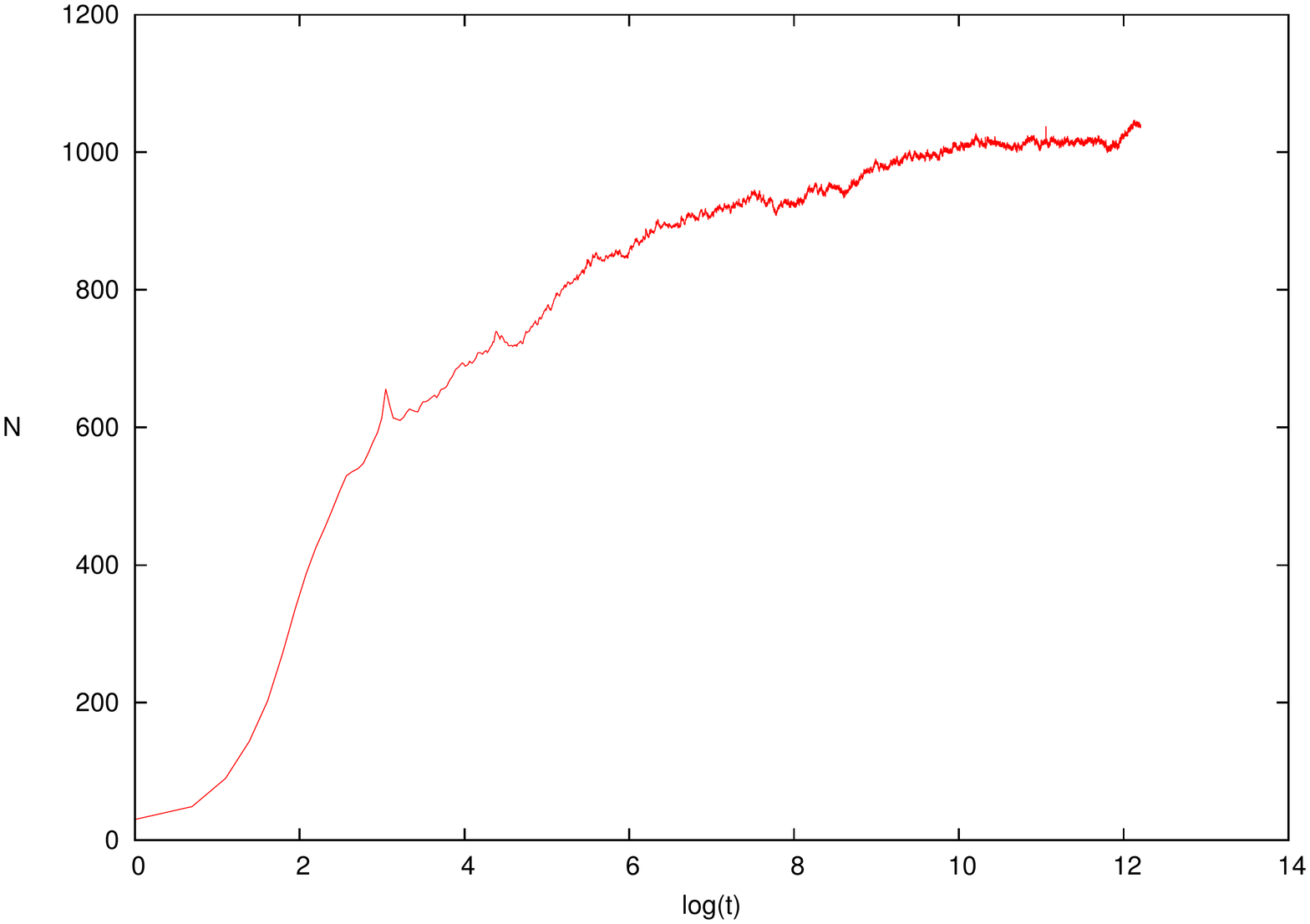}
        \caption{ Total population, $N$, averaged over 1000 runs, $\sigma=0.1$. }
        \label{fig:N}
    \end{subfigure}  \qquad
    \begin{subfigure}[t]{0.4\textwidth}
        \includegraphics[width=\textwidth]{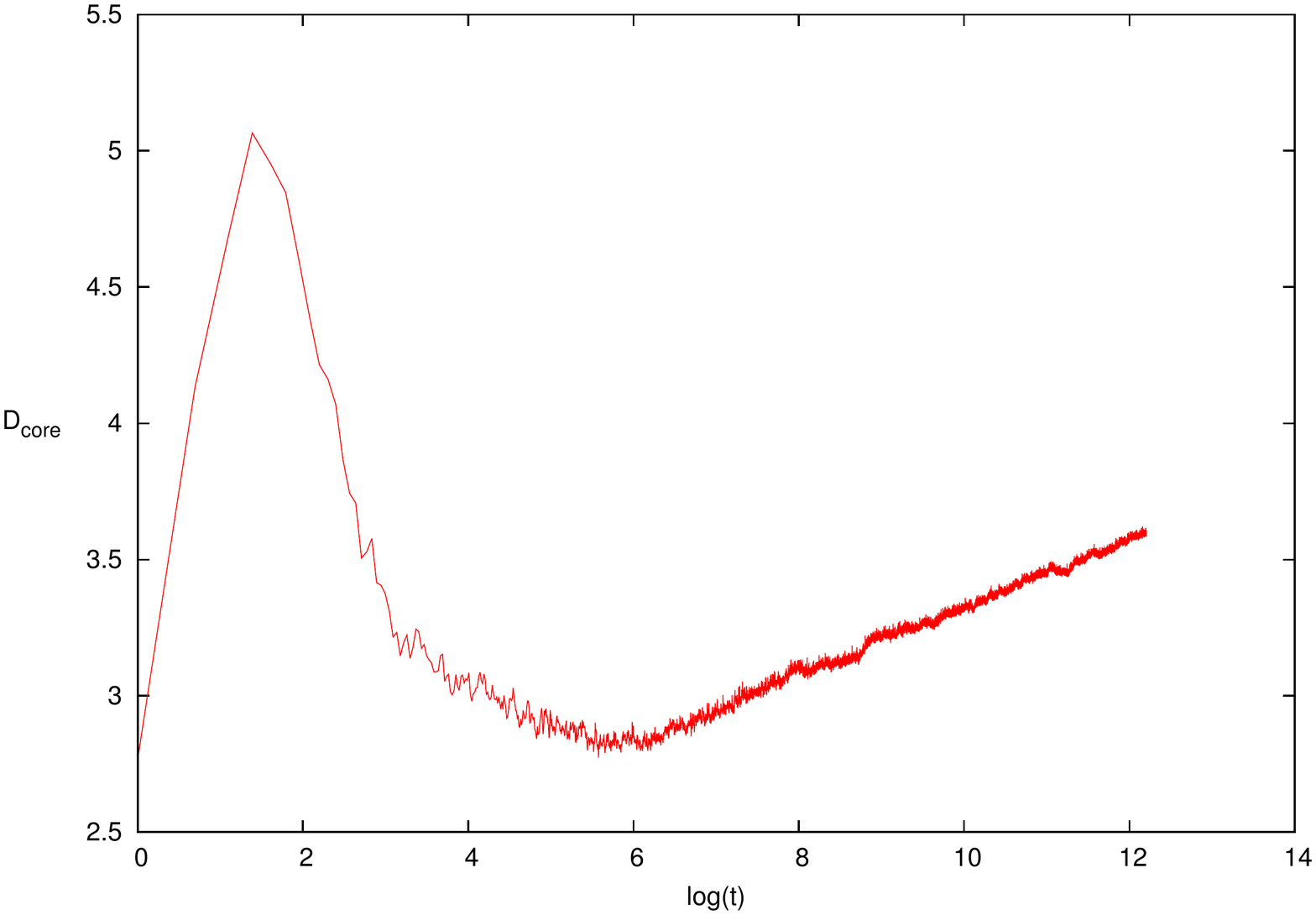}
        \caption{ Number of core species $D_{core}$, averaged over 1000 runs, $\sigma=0.1$. }
        \label{fig:Nc}
    \end{subfigure}  
    
    \begin{subfigure}[t]{0.4\textwidth}
        \includegraphics[width=\textwidth]{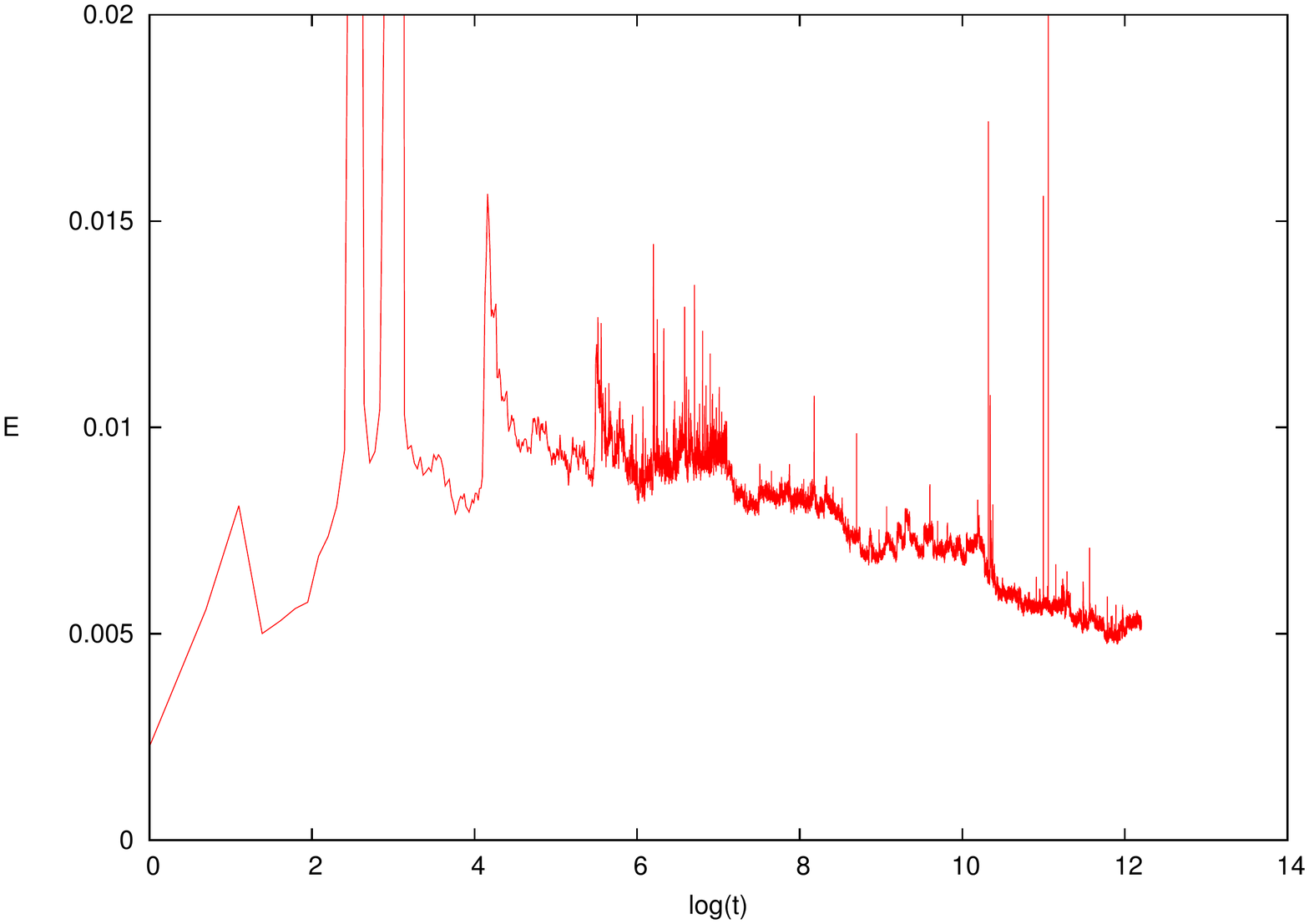}
        \caption{ Habitability $E$, averaged over 1000 runs, $\sigma=0.1$. }
        \label{fig:E}
    \end{subfigure} \qquad
        \begin{subfigure}[t]{0.4\textwidth}
        \includegraphics[width=\textwidth]{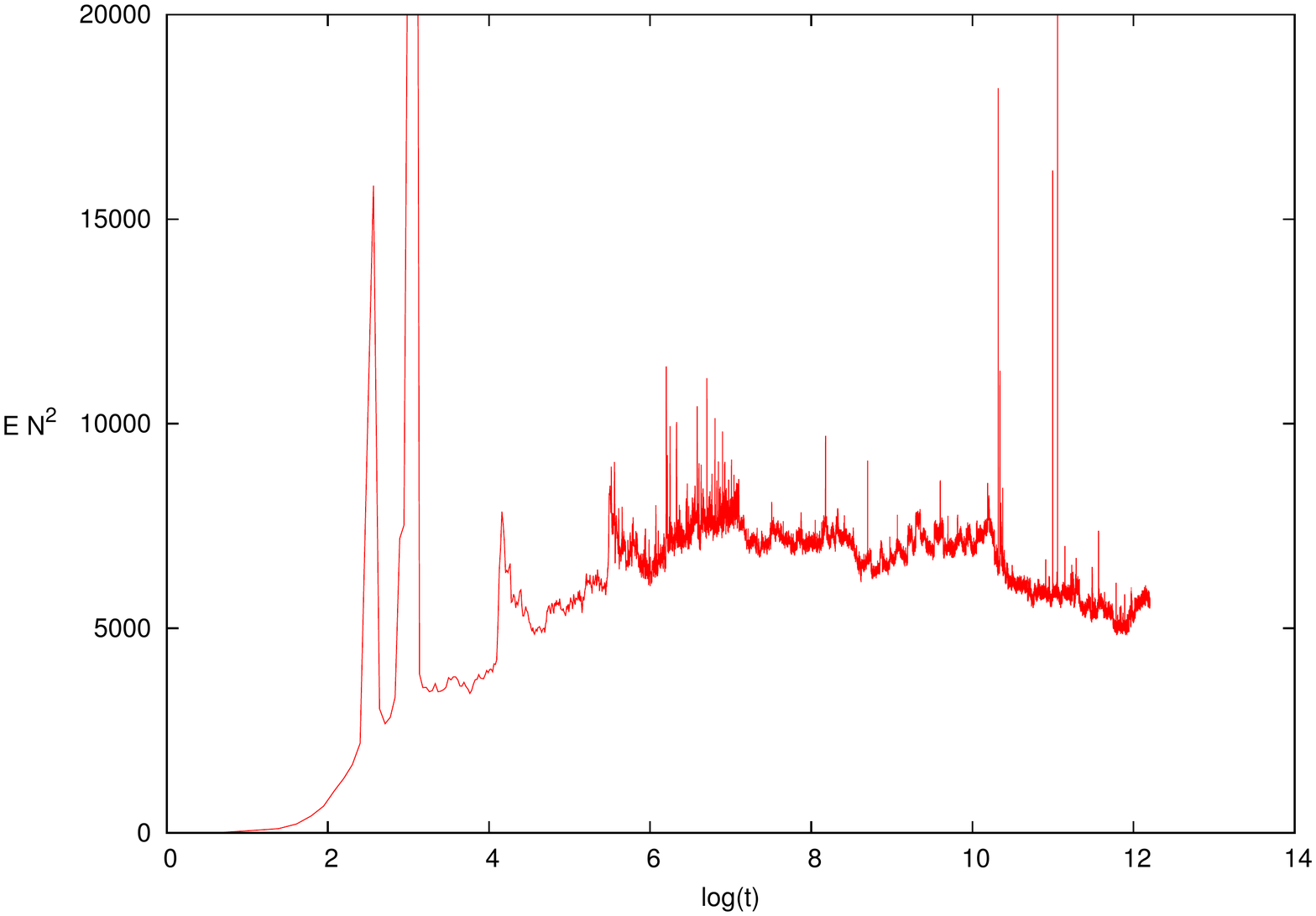}
        \caption{ Total environmental effect $E N^2$, averaged over 1000 runs, $\sigma=0.1$. }
        \label{fig:ENN}
    \end{subfigure} 
    \caption{Averages of TNM observables across 1000 realisations of the model. Plotted against the natural log
    of the generation number $t$. }\label{fig:av}
\end{figure}

In figure \ref{fig:av} we show the average of several key observables over an ensemble of 1000 different
realisations of the model. Total population and the number of species in the core increases logarithmically 
as in the standard TNM. $E$ is positive with a `n' shape, as is the total effect of all individuals on the environment, $EN^2$, 
figure \ref{fig:ENN}. This means that the species that are able to survive and reproduce in the TNM tend to
improve the environment, though the strength of this effect changes over time. We also note some very large
fluctuations in $E$. These are caused by runs where a very strong positive environmental feedback
was established between two species, as in equation \ref{eqn:explosion}, but before it could run away a mutant arose which
stopped the exponential population growth.

\section{Discussion}\label{sec:discussion}

The reasons for the gradual increase in TNM population show one possible way that evolution
by natural selection can lead to better conditions for life. 
An individual run of the TNM is usually in a quasi-stable state with a core
consisting of $D_{core}$ core species and $N_{core}$ total individuals. Mutations of the core create a cloud of $N_{cloud}$ 
mutants of $D_{cloud}$ different species. $N_{core} \gg N_{cloud}$ and $D_{cloud} \gg D_{core}$.
In a quasi-stable state the reproduction and death probabilities are roughly equal for each core species
$$
p(f_c) \simeq p_k \implies f_c = \sum_{j} J_{cj}^{eff} n_j - \mu N \simeq \log\left( \frac{p_k}{1-p_k} \right)
$$ 
A stable state ends when a new species $a$ arises which has a significant reproduction probability
$p(f_a) > p_k$. Because of the damping term $-\mu N$ this means the new species needs strong positive interactions,
$J_{ac}^{eff} \gg 0$, with species in the core. This new species will grow exponentially since
$$
\frac{dN_a}{dt} \simeq N_a (p(f(a)) - p_k) > 0.
$$
The new species has a negative effect on the core through the $-\mu N$ term, by which all species are coupled,
and can also have some `parasitic' couplings $J_{ca}^{eff} < 0$. Even if the couplings are positive,
if they are too small to compensate for the change in the $-\mu N$ term
$f_c$ will be reduced making $p(f_c) < p_k$, so the population of the core species will exponentially decrease.
Since large populations of all core species are necessary to support each other, this will
result in the collapse of the core, as well as the species $a$. The result is
a partial vacuum of many sparsely populated species, members of the old cloud and
remnants of the core. We will call this a `parasitic quake'.
Another mechanism occurs in cases where the new species has `symbiotic' interactions ($J^{eff}_{ac} > 0$, $J^{eff}_{ca} > 0$)
with all the core species that are strong enough to maintain $p(f_c) > p_k$. Instead of a core collapse this causes
a core rearrangement, where the new species is incorporated into the core and the relative populations of each species change.
This is a `symbiotic quake'.

\begin{figure}
    \centering
    \begin{subfigure}[t]{0.4\textwidth}
        \includegraphics[width=\textwidth]{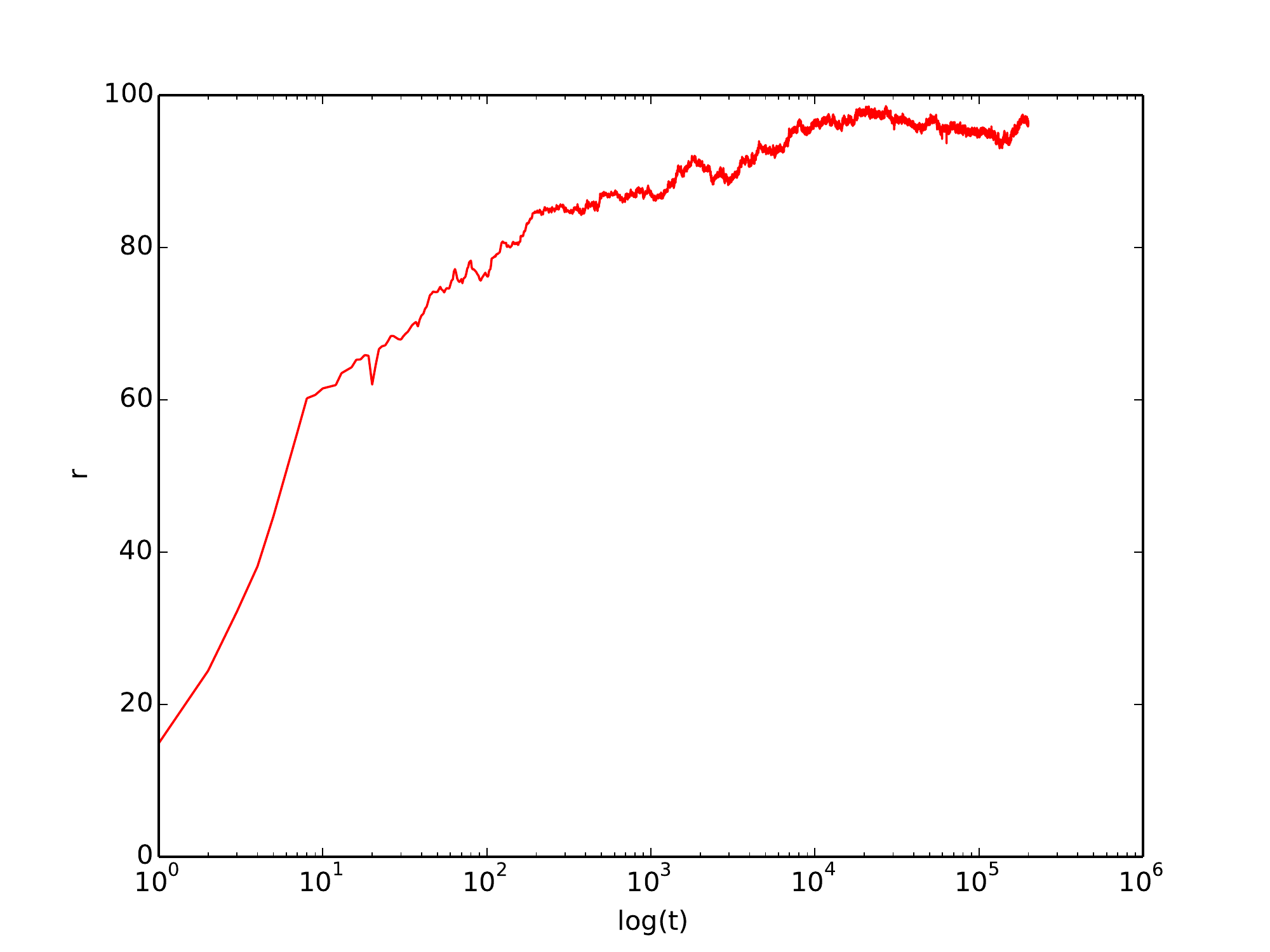}
        \caption{ Average growth rate $r$, $\sigma=0.1$. }
        \label{fig:r}
    \end{subfigure} \qquad
        \begin{subfigure}[t]{0.4\textwidth}
        \includegraphics[width=\textwidth]{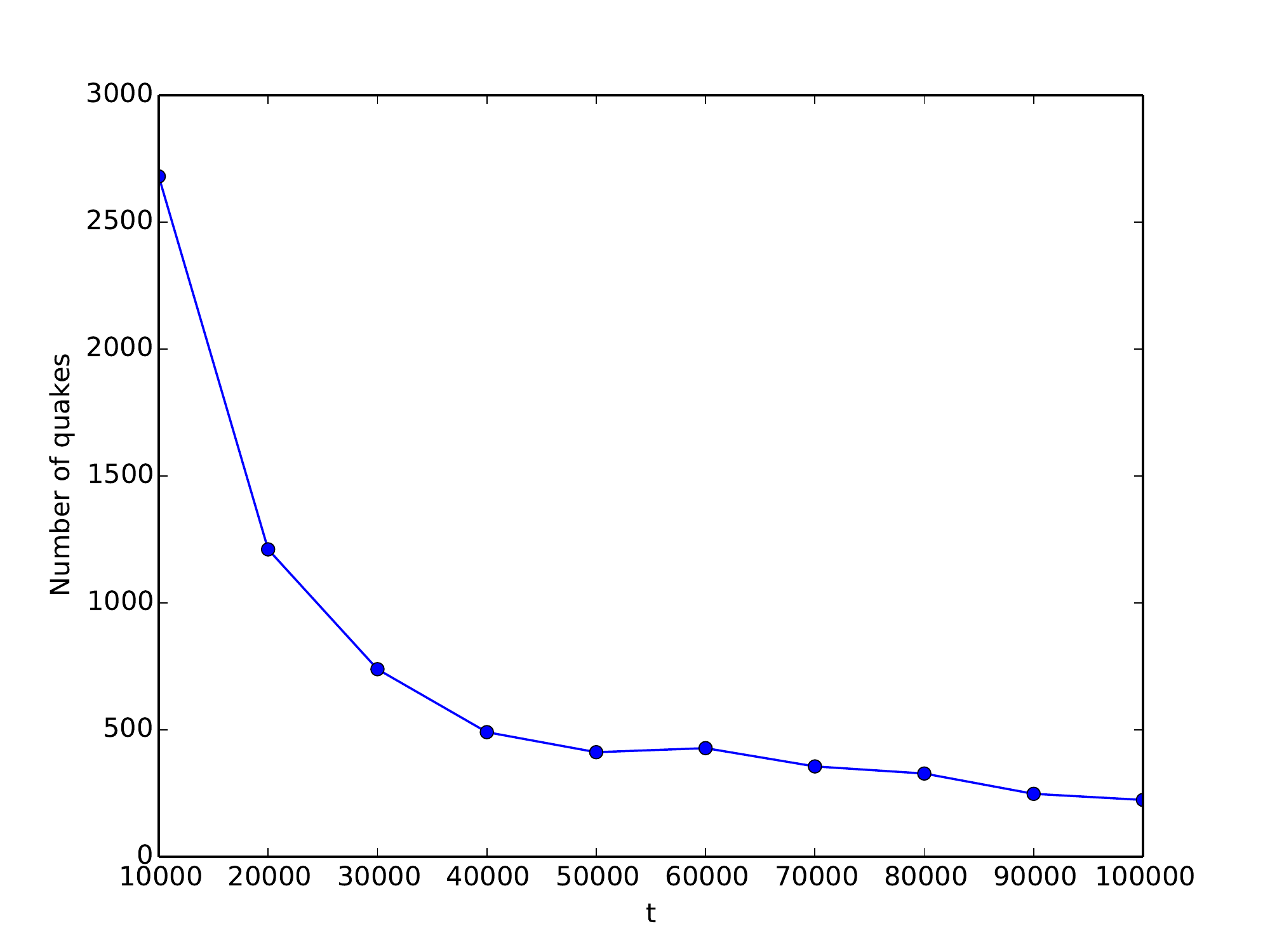}
        \caption{ Number of quakes per 10000 generations. }
        \label{fig:Q}
    \end{subfigure} 
    \caption{(a) Average growth rate $r = \sum_{ij} n_i J_{ij} n_j$ as a function of time.
    (b) We operationally define a stable period as 100 generations during which the species in the core are unchanged.
    The number of quakes is then one minus the number of stable periods. The plot shows the number of quake events
    occurring in ten 10000-generation windows. }\label{fig:stab}
\end{figure}

After a parasitic quake the core which arises is the one with the highest reproduction probability per species that can be
formed from previously existing cloud species or genetically close mutants. Every quake is a trial where the system `chooses'
the strongest group from a set of possible species. The most optimal core is not necessarily chosen, but it is
more likely to be. This repeated selection
gradually increases the average reproduction probability and hence the average population, as observed in
figure \ref{fig:r}. This also leads to an increase in stability with time. For a species $a$ to destabilize or join the core
it has to have sufficiently large reproduction probability (see \cite{Becker:2013} for much more discussion
of this point),
$$
p(f_a) > p_k \implies \sum_j J_{aj}^{eff} n_j > \log\left( \frac{p_k}{1-p_k} \right) + \mu N
$$
Since $N$ is gradually increasing this represents a gradually increasing barrier for parasites to be viable and hence longer
stable periods, figure \ref{fig:Q}.

Looking at how quakes occur explains the weaker selection for improving the environment i.e. for negative values of
$M_{ij}$. Simply, the term $M_{ij} N_j$ is small for new species ($N_j=1$) and doesn't affect the reproduction
until $N_j$ becomes large. We still do have some selection for smaller $M_{ij}$ in the
following way: if a new core starts to grow rapidly, but one or more of the $M_{ij}$ terms is large and positive,
as $N_j$ increases the growth of species $i$ slows. This can enable a different potential core, without a
limiting $M_{ij}$, to overtake and dominate. As long as the values of $M_{ij}$ are small (or negative)
enough to allow the potential core, selected on the basis of the $J_{ij}$s, to grow and dominate the ecology
then then they will be observed. 

   \begin{figure}[t]
   \centering
        \includegraphics[width=\textwidth]{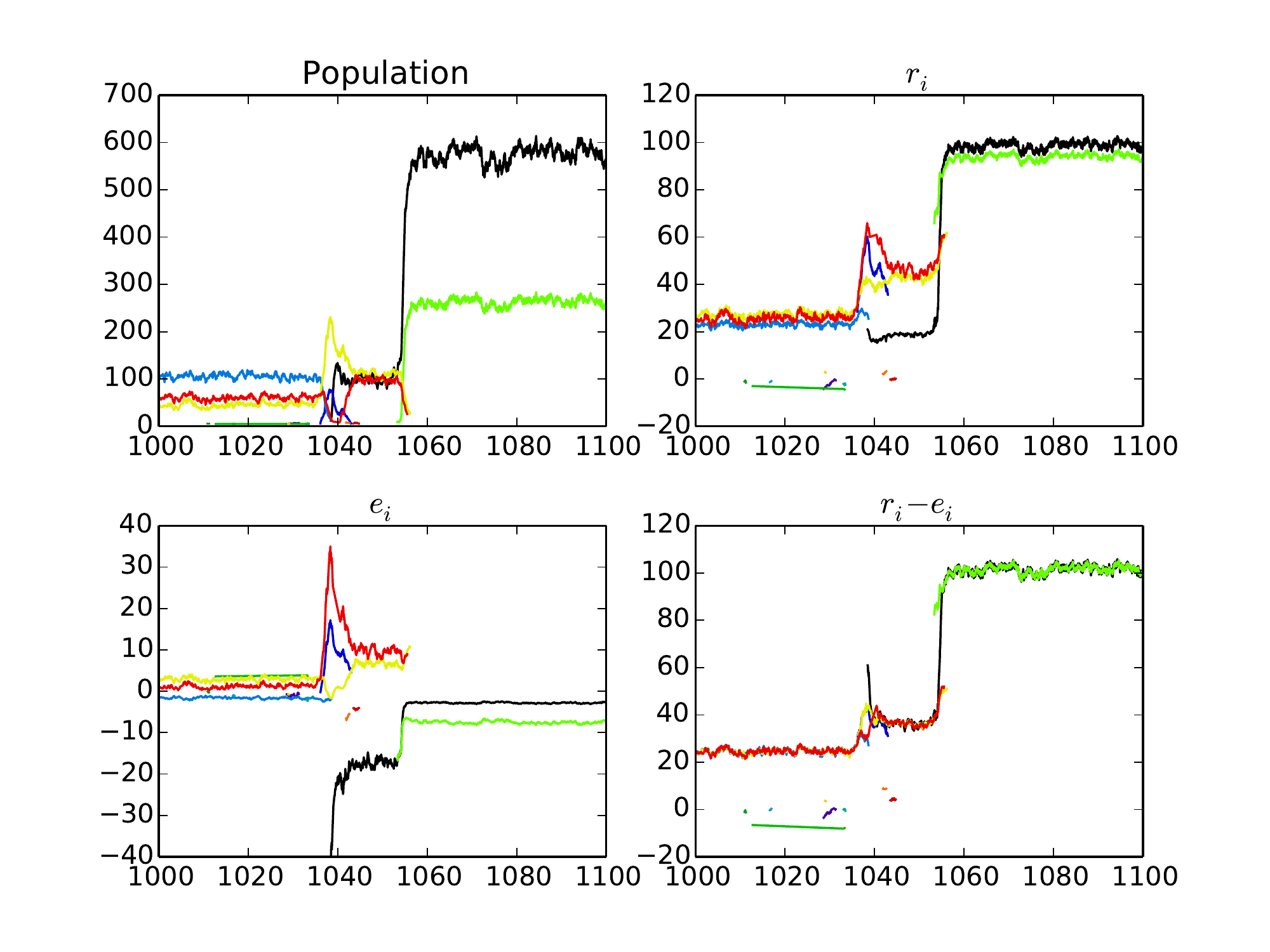}
        \caption{ High resolution plot of population (top left), $r_i$ (top right), $e_i$ (bottom left) and 
        growth rate $r_i-e_i$ (bottom right) for the
        most populous species in a single TNM run just after a quake. Different colours represent different species. 
        x-axis measures generation number. Colour online.}
        \label{fig:fine}
    \end{figure}  
    
We can see this by examining a quake in detail, figure \ref{fig:fine}. We have taken a
snapshot of a system just before and after a quake. During the quake,
between 1040 and 1060, we have a new potential core of yellow, red and black. The plot of $r_i$ shows this is 
positive all for three, so they have mutually symbiotic interactions. But $e_i$ for red and yellow is large and
positive, so the environment created by these three species is not beneficial for two of them and the net growth rate
$r_i - e_i$ is smaller than otherwise. Thus when the
green species (present in low numbers, or a nearby mutant of one of the more populous species) starts to reproduce, 
with a large net growth rate $r_i-e_i$, it quickly establishes itself as part of a new, long lived, stable core.
The values of $J_{ij}$ are positive for all mutual interactions between red, yellow, black and green. Thus the green species
is able to establish itself without parasitising one of the other species, simply its very large growth rate
allows it to overtake the others and `use up' the $\mu N$ term, so that reproduction for other species is unlikely.
If the signs of $e_i$ were reversed for the red and yellow species, they could have saturated the $-\mu N$ term themselves,
preventing the rise of the green species. Cores with negative effects on the environment can and do establish themselves, 
however they are more fragile since they have smaller populations, and hence smaller barriers for new species to overcome. 

The tendency towards higher $r_i$ values means the effect of
the environment becomes less important, especially during the key period when the new core is exponentially growing.
The new core can grow large before $e_i$ becomes significant 
and so is less likely to be disrupted. However since the growth rate increases 
very slowly (logarithmically) the average contribution of an individual to the habitability is 
still positive even after $2 \times 10^5$ generations, so $E$ remains significant for a long time.

%boxes
Larger populations mean more mutants in the cloud so, if there was a quake, there would be many more
potential configurations to choose from and hence larger populations likely to occur in the new core. 
However quakes become less likely with time! This process is an example of an entropic hierarchy. 
A simple example of an entropic hierarchy is a stack of boxes of increasing size with a single small 
hole connecting each box to the one below and the one above. A particle bounces around a small box until it finds
a hole and escapes.
If the particle escapes back to a smaller box it exits again quickly, but if it goes to the larger box then it stays there
longer. Finding a hole is an example of an entropic barrier. The particle is most likely to be in the largest box 
(where the number of possible configurations is largest), 
but may be trapped in a smaller box for a significant time. The entropic barrier in the TNM is the difficulty of
generating a mutant with sufficiently large interactions to destabilize the core.
Once such a mutant is found the system moves into a new configuration space, with more potential cores to choose from
i.e. more configurations and higher entropy! Because a strongly symbiotic core is likely to be realised 
a new mutant needs an even larger interaction strength to overcome the next barrier. Finding a species with an interaction
large enough to overcome this barrier becomes less and less likely 
because of how the $J_{ij}$s are distributed, with larger values being exponentially rarer.

\section{Conclusion}\label{sec:conclusion}

The TNM is closely related to the Logistic model of population dynamics, but incorporates co-evolution by
making the reproduction rate depend on the other species present. We have extended this by allowing the damping term
in the Logistic model to depend on the species that are present - so that they can affect the amount of resources available
in each other's environments. The phenomenology of the standard TNM is reproduced by this model 
and we also find that systems evolve on average so that species' effect on each other's resources is positive.
Following \cite{Becker:2013}, we showed increasing stability is due to increasing entropic barriers - namely
the increasing difficulty of creating a destabilizing mutant. 
This makes it more likely that mutually beneficial systems, with high populations, persist, since the height of 
the barrier increases with population. 
Quasi-stable configurations are mostly selected for on the basis of their direct 
interactions, but if some species reduce habitability this reduces the total population and makes 
the system less stable than if those species improved it. Thus periods with environment degrading species 
are shorter than periods with environment improving species. This process has been described as `sequential selection' 
\cite{Betts:2007}, \cite{Dolittle:2014}.

Some of controversy over Gaia has been due to trying to explain Gaian effects in terms of natural selection.
As has been correctly argued \cite{Tyrrell:2013}, 
this cannot account for the development of positive environmental feedbacks which
don't directly benefit an individual. However Gaian ideas of species-environment co-evolution leading
to increases in stability and habitability are not contingent on this. The model described in this work 
leads to the conclusions that, when species and environment co-evolve: stability increases as a consequence of 
increasing entropic barriers and habitability is positively affected by life due to sequential selection.
Note that life does not {\it necessarily} improve the environment in 
this model, but we can make statements  about averages over ensembles of possible realisations of a system's history. 

Daisyworld type models \cite{Watson:1983} look at the effect of external perturbations on a
simplified earth system (particularly the effect of increasing solar luminosity with time).
This model only has internal perturbations, but it is possible to allow for
non-biologically driven external perturbations, by letting $\mu$ to vary with time (see \cite{Arthur:2016}
for example). This is an interesting future direction, but the main message of this, and many other agent based
models, is that ecologies are capable of endogenously generating catastrophes. Our work
shows how the ecologies adapt to reduce the frequency of these with time, while improving their
capacity to support life, by climbing an entropic hierarchy. Gaia - meaning stability and
habitability - can arise from entropy.

\section*{Acknowledgements}
The authors would like to thank Tim Lenton and Hywel Williams for their comments on this manuscript.

\end{document}